\documentclass[a4paper]{article}
\RequirePackage[english]{babel}
\RequirePackage[latin1]{inputenc}
\RequirePackage[T1]{fontenc}
\RequirePackage{mathrsfs}
\RequirePackage{amsmath}
\RequirePackage{amssymb}
\RequirePackage{amsbsy}
\RequirePackage{bm}
\begin{document}
\title{\bf{Zero Energy of Plane-Waves for ELKOs}}
\author{Luca Fabbri}
\date{}
\maketitle
\begin{abstract}
We consider the ELKO field in interaction through contorsion with its own spin density, and we investigate the form of the consequent autointeractions; to do so we take into account the high-density limit and find plane wave solutions: such plane waves give rise to contorsional autointeractions for which the Ricci metric curvature vanishes and therefore the energy density is equal to zero identically. Consequences are discussed.
\end{abstract}
\section*{Introduction}
Recently, Ahluwalia and Grumiller have defined a new form of matter, which they called ELKOs, where the name is the acronym of the German that stands for eigenspinors of the charge-conjugation operator, as they are defined to be spin-$\frac{1}{2}$ spinors autoconjugated under the charge conjugation operator; these spin-$\frac{1}{2}$ spinors have two irreducible projections that are two semi-spinors charge-conjugated to one another: roughly speaking an ELKO has the same degrees of freedom of any spin-$\frac{1}{2}$ spinor but rearranged in such a way that the result is actually a topologically neutral spinor field.

Starting from this definition, it is possible to see that the most fundamental property of ELKOs is that they turn out to have mass dimension equal to $1$ so that their dynamics has to be constructed upon scalar-like field equations; thus the ELKO, being a spinor, has derivatives that contain the spacetime contorsion, and, having scalar-like field equations, it has second-order derivatives in its field equations: so a richer dynamics arises (\cite{a-g/1}, \cite{a-g/2}).

One of the most important problems that this situation creates is that contorsion is related to the spin of the field which is written in terms of the derivatives containing contorsion; this is a problem because in this way contorsion is defined through an implicit relationship which must be inverted if we want the expression of contorsion to be written in terms of the contorsionless derivatives of the field alone: in recent papers this has been done and thus the explicit contorsion has been obtained. Hence it has now become possible to employ contorsion directly in all equations to see what are all contorsional effects in the dynamics of the ELKO fields. For instance, one of these contorsional effects is that, for such second-order derivative field equations, there are derivatives of the contorsion, which is itself written in terms of the derivatives of the field, and thus derivatives of derivatives of the field appear beside the usual D'Alembertian of the field; this contorsional effects gives rise to back-reactions that may create the problem of causal propagation: however in this specific case, the problem of the causal propagation has been solved affirmatively (\cite{f/1} and \cite{f/2}).

The very definition of ELKO and their dynamics are the basis for the explanation of some fundamental problems in cosmology and hence the reason for which ELKOs are so important for cosmology's standard model: the idea of a privileged direction arising from a preferred axis (\cite{l-m}, \cite{f-e}) could be explained by their spin structure (\cite{a-l-s/1}, \cite{a-l-s/2}); the curves of rotation of galaxies and the inflationary expansion of the universe (\cite{s-s}, \cite{a-s}, \cite{b-s-i}) could be described by their dynamics (\cite{a}, \cite{b-m}, \cite{b/2}, \cite{b/11}, \cite{s/1}, \cite{s/2}). Some of these results have been reviewed while others have been extended and a comprehensive list of the results is accounted in the literature (\cite{a-h}, \cite{r-r}, \cite{r-h}, \cite{h-r}, \cite{b-b/2}, \cite{b-b/1}, \cite{b/1}, \cite{b-b}, \cite{b-b-m-s}).

In the present paper we consider the explicit form of contorsion directly written in all equations to investigate some of the consequences of the contorsional contributions for the ELKO fields.
\section{The ELKO Field Theory}
In this paper the spacetime connection will be given by $\Gamma^{\mu}_{\alpha\sigma}$ and it will be used to define the Riemann curvature tensor as
\begin{eqnarray}
G^{\rho}_{\phantom{\rho}\eta\mu\nu}
=\partial_{\mu}\Gamma^{\rho}_{\eta\nu}-\partial_{\nu}\Gamma^{\rho}_{\eta\mu}
+\Gamma^{\rho}_{\sigma\mu}\Gamma^{\sigma}_{\eta\nu}
-\Gamma^{\rho}_{\sigma\nu}\Gamma^{\sigma}_{\eta\mu}
\end{eqnarray}
which has one independent contraction given by $G^{\rho}_{\phantom{\rho}\eta\rho\nu}=G_{\eta\nu}$ whose contraction is given by $G_{\eta\nu}g^{\eta\nu}=G$ as usual and then we define Cartan torsion tensor as
\begin{eqnarray}
Q^{\rho}_{\phantom{\rho}\mu\nu}
=\Gamma^{\rho}_{\mu\nu}-\Gamma^{\rho}_{\nu\mu}
\end{eqnarray}
and contorsion tensor
\begin{eqnarray}
K^{\rho}_{\phantom{\rho}\mu\nu}-K^{\rho}_{\phantom{\rho}\nu\mu}
=\Gamma^{\rho}_{\mu\nu}-\Gamma^{\rho}_{\nu\mu}
\end{eqnarray}
so that torsion and contorsion are linked by the relationship
\begin{eqnarray}
K^{\rho}_{\phantom{\rho}\mu\nu}
=\frac{1}{2}\left(Q^{\rho}_{\phantom{\rho}\mu\nu}
+Q_{\mu\nu}^{\phantom{\mu\nu}\rho}+Q_{\nu\mu}^{\phantom{\nu\mu}\rho}\right)
\end{eqnarray}
with one independent contraction given by $K_{\nu\rho}^{\phantom{\rho\nu}\rho}= Q^{\rho}_{\phantom{\rho}\rho\nu}=Q_{\nu}=K_{\nu}$ as convention: when in the connection the torsion or contorsion tensors are separated away we are left with the symmetric connection in terms of which we define the Riemann metric curvature tensor $R^{\rho}_{\phantom{\rho}\eta\mu\nu}$ with one contraction $R_{\eta\nu}$ whose contraction is $R$ as in the usual geometry. As a final definition we have that the contraction of the Riemann curvature tensor will be called Ricci curvature tensors: then the contractions of the Riemann metric curvature tensor will be called Ricci metric curvature tensors. By using these tensors we can build the relationship between the curvature and the energy density $T^{\mu\nu}$ as well as the relation between the contorsion and the spin density $S^{\rho\mu\nu}$ in terms of the Einstein-Sciama-Kibble field equations given by
\begin{eqnarray}
G^{\mu\nu}-\frac{1}{2}g^{\mu\nu}G=\frac{1}{2}T^{\mu\nu}
\label{energy}
\end{eqnarray}
and
\begin{eqnarray}
\left(K_{\mu[\alpha\beta]}+K_{[\alpha}g_{\beta]\mu}\right)=-S_{\mu\alpha\beta}
\label{spin}
\end{eqnarray}
where the gravitational constant has been normalized: again it is possible to separate contorsion everywhere in the equations for the energy (\ref{energy}) and to invert contorsion in terms of the spin in the equations for the spin (\ref{spin}) in order to substitute it into the field equations for the energy (\ref{energy}) obtaining the field equations that relate the metric curvature tensor to the energy density. These field equations give a combination of the Ricci metric curvature tensor in terms of the energy density: therefore we need to invert them if we want to obtain field equations relating the Ricci metric curvature tensor to a combination of the energy density. The matter fields we will employ will be spin-$\frac{1}{2}$ fermion fields defined in terms of the transformation law $\phi'=S\phi$ where $S$ is a complex representation of the Lorentz group and for which the derivatives $D_{\mu}$ are defined by the contorsionless derivatives $\nabla_{\mu}$ according to the decomposition 
\begin{eqnarray}
D_{\mu}\lambda=\nabla_{\mu}\lambda+\frac{1}{2}K^{ij}_{\phantom{ij}\mu}\sigma_{ij}\lambda
\end{eqnarray}
in terms of the $\sigma_{ij}$ matrices. The commutator of the derivatives is given by
\begin{eqnarray}
[D_{\mu},D_{\nu}]\lambda=Q^{\rho}_{\phantom{\rho}\mu\nu}D_{\rho}\lambda
+\frac{1}{2}G^{ij}_{\phantom{ij}\mu\nu}\sigma_{ij}\lambda
\end{eqnarray}
in terms of the $\sigma_{ij}$ matrices given as $\sigma_{ij}=\frac{1}{4}[\gamma_{i},\gamma_{j}]$ where the gamma matrices are to satisfy the anticommutation relationships represented by the Clifford algebra; from them we also define $i\gamma^{0}\gamma^{1}\gamma^{2}\gamma^{3}=\gamma$ as the gamma pseudo-matrix.

For the Ahluwalia-Grumiller ELKO field and its dual we will use the notation $\lambda$ and $\stackrel{\neg}{\lambda}$ defined in the above references; we shall now give the field equations for the ELKO and the ELKO dual that will be used to obtain from the conserved quantities given by the energy and spin density the conservation laws for which the Einstein-Sciama-Kibble field equations will be satisfied.

The ELKO matter field equations are given by 
\begin{eqnarray}
D^{2}\lambda+K^{\mu}D_{\mu}\lambda+m^{2}\lambda=0
\end{eqnarray}
and its dual, where $m$ is the mass and in which in the derivatives contorsion can be separated leaving contorsionless derivatives (\cite{a-g/1} and \cite{a-g/2}).

The conserved quantities are given by the energy
\begin{eqnarray}
T_{\mu\nu}=
\left(D_{\mu}\stackrel{\neg}{\lambda}D_{\nu}\lambda+D_{\nu}\stackrel{\neg}{\lambda}D_{\mu}\lambda
-g_{\mu\nu}D_{\rho}\stackrel{\neg}{\lambda}D^{\rho}\lambda
+g_{\mu\nu}m^{2}\stackrel{\neg}{\lambda}\lambda\right)
\end{eqnarray}
and the spin
\begin{equation}
S_{\mu\alpha\beta}=\frac{1}{2}\left(D_{\mu}\stackrel{\neg}{\lambda}\sigma_{\alpha\beta}\lambda
-\stackrel{\neg}{\lambda}\sigma_{\alpha\beta}D_{\mu}\lambda\right)
\end{equation}
in which the spin is written in terms of the spinor field derivatives containing contorsion and thus determining contorsion as an implicit relationship which can be inverted to explicitly give contorsion in terms of the contorsionless derivatives of the spinor fields (\cite{f/1} and \cite{f/2}).

After the contorsion is separated we explicitly get the field equations
\begin{eqnarray}
\nonumber
&\nabla^{2}\lambda+K^{ij\mu}\sigma_{ij}\nabla_{\mu}\lambda
+\frac{1}{2}\nabla_{\mu}K^{\alpha\beta\mu}\sigma_{\alpha\beta}\lambda-\\
&-\frac{1}{8}K^{ij\mu}K_{ij\mu}\lambda
+\frac{i}{16}K_{ij}^{\phantom{ij}\mu}K_{ab\mu}\varepsilon^{ijab}\gamma\lambda+m^{2}\lambda=0
\end{eqnarray}
in terms of the contorsion tensor.

And after contorsion is inverted its explicit form is given by
\begin{eqnarray}
\nonumber
&K_{\alpha\beta\mu}
\left[(8+\stackrel{\neg}{\lambda}\lambda)^{2}
+(i\stackrel{\neg}{\lambda}\gamma\lambda)^{2}\right]
\left[(4-\stackrel{\neg}{\lambda}\lambda)^{2}
+(i\stackrel{\neg}{\lambda}\gamma\lambda)^{2}\right]=\\
\nonumber
&=\left[(8+\stackrel{\neg}{\lambda}\lambda)^{2}
+(i\stackrel{\neg}{\lambda}\gamma\lambda)^{2}\right]
(i\stackrel{\neg}{\lambda}\gamma\lambda)
(\stackrel{\neg}{\lambda}\sigma^{\sigma\rho}\nabla_{\mu}\lambda
-\nabla_{\mu}\stackrel{\neg}{\lambda}\sigma^{\sigma\rho}\lambda)
\varepsilon_{\sigma\rho\alpha\beta}-\\
\nonumber
&-2\left[(8+\stackrel{\neg}{\lambda}\lambda)^{2}
+(i\stackrel{\neg}{\lambda}\gamma\lambda)^{2}\right]
(4-\stackrel{\neg}{\lambda}\lambda)
(\stackrel{\neg}{\lambda}\sigma_{\alpha\beta}\nabla_{\mu}\lambda
-\nabla_{\mu}\stackrel{\neg}{\lambda}\sigma_{\alpha\beta}\lambda)-\\
\nonumber
&-4\left[(4-\stackrel{\neg}{\lambda}\lambda)
(8+\stackrel{\neg}{\lambda}\lambda)
+(i\stackrel{\neg}{\lambda}\gamma\lambda)^{2}\right]
(\stackrel{\neg}{\lambda}\sigma_{\sigma\theta}\nabla_{\zeta}\lambda
-\nabla_{\zeta}\stackrel{\neg}{\lambda}\sigma_{\sigma\theta}\lambda)
\varepsilon_{\alpha\beta\mu\rho}\varepsilon^{\sigma\theta\zeta\rho}-\\
\nonumber
&-8\left[(4-\stackrel{\neg}{\lambda}\lambda)
(8+\stackrel{\neg}{\lambda}\lambda)
+(i\stackrel{\neg}{\lambda}\gamma\lambda)^{2}\right]
(\stackrel{\neg}{\lambda}\sigma_{\eta\alpha}\nabla^{\eta}\lambda
-\nabla^{\eta}\stackrel{\neg}{\lambda}\sigma_{\eta\alpha}\lambda)g_{\mu\beta}+\\
\nonumber
&+8\left[(4-\stackrel{\neg}{\lambda}\lambda)
(8+\stackrel{\neg}{\lambda}\lambda)
+(i\stackrel{\neg}{\lambda}\gamma\lambda)^{2}\right]
(\stackrel{\neg}{\lambda}\sigma_{\eta\beta}\nabla^{\eta}\lambda
-\nabla^{\eta}\stackrel{\neg}{\lambda}\sigma_{\eta\beta}\lambda)g_{\mu\alpha}-\\
\nonumber
&-16(2+\stackrel{\neg}{\lambda}\lambda)
(i\stackrel{\neg}{\lambda}\gamma\lambda)
(\stackrel{\neg}{\lambda}\sigma^{\eta\rho}\nabla_{\eta}\lambda
-\nabla_{\eta}\stackrel{\neg}{\lambda}\sigma^{\eta\rho}\lambda)\varepsilon_{\alpha\beta\mu\rho}+\\
\nonumber
&+8(2+\stackrel{\neg}{\lambda}\lambda)
(i\stackrel{\neg}{\lambda}\gamma\lambda)
(\stackrel{\neg}{\lambda}\sigma^{\sigma\theta}\nabla^{\zeta}\lambda
-\nabla^{\zeta}\stackrel{\neg}{\lambda}\sigma^{\sigma\theta}\lambda)
g_{\mu\beta}\varepsilon_{\sigma\theta\zeta\alpha}-\\
&-8(2+\stackrel{\neg}{\lambda}\lambda)
(i\stackrel{\neg}{\lambda}\gamma\lambda)
(\stackrel{\neg}{\lambda}\sigma^{\sigma\theta}\nabla^{\zeta}\lambda
-\nabla^{\zeta}\stackrel{\neg}{\lambda}\sigma^{\sigma\theta}\lambda)
g_{\mu\alpha}\varepsilon_{\sigma\theta\zeta\beta}
\end{eqnarray}
in terms of the contorsionless derivatives of the matter field which can be used to invert the metric curvature field equations to give the Ricci metric curvature tensor as
\begin{eqnarray}
\nonumber
&R_{\mu\nu}=\frac{1}{2}\left(\nabla_{\mu}\stackrel{\neg}{\lambda}\nabla_{\nu}\lambda
+\nabla_{\nu}\stackrel{\neg}{\lambda}\nabla_{\mu}\lambda
-g_{\mu\nu}m^{2}\stackrel{\neg}{\lambda}\lambda\right)-\\
\nonumber
&-\frac{1}{8}K^{\sigma\rho}_{\phantom{\sigma\rho}\mu}K_{\sigma\rho\nu}
\left(\stackrel{\neg}{\lambda}\lambda\right)
+\frac{1}{16}K_{\sigma\rho\mu}K_{\eta\zeta\nu}\varepsilon^{\sigma\rho\eta\zeta}
\left(i\stackrel{\neg}{\lambda}\gamma\lambda\right)-\\
\nonumber
&-\frac{1}{2}\nabla^{\rho}\left(K_{\rho\mu\nu}+K_{\rho\nu\mu}
+g_{\rho\nu}K_{\mu}+g_{\rho\mu}K_{\nu}\right)-\\
&-\frac{1}{2}\left(K^{\sigma\rho}_{\phantom{\sigma\rho}\nu}K_{\mu\sigma\rho}
+K^{\sigma\rho}_{\phantom{\sigma\rho}\mu}K_{\nu\sigma\rho}
+K^{\rho}K_{\rho\mu\nu}+K^{\rho}K_{\rho\nu\mu}\right)
\end{eqnarray}
in terms of the contorsionless derivatives of the matter fields themselves.

\subsection{Plane-wave solutions, vanishing of the energy density}
In order to deepen the previous analysis we shall look for what we believe to be the most interesting limit, that is the high-density approximation given when the bilinear of the field tend to have values that, after a suitable normalization is chosen, they are very high, compared to the unity; furthermore, we will assume that within the derivative terms we will have the presence of the contorsion whereas the contorsionless metric connection are negligible: in this approximation the contorsion tensor is approximated to
\begin{eqnarray}
\nonumber
&K_{\alpha\beta\mu}\left[(\stackrel{\neg}{\lambda}\lambda)^{2}
+(i\stackrel{\neg}{\lambda}\gamma\lambda)^{2}\right]=\\
\nonumber
&=(i\stackrel{\neg}{\lambda}\gamma\lambda)
(\stackrel{\neg}{\lambda}\sigma^{\sigma\rho}\nabla_{\mu}\lambda
-\nabla_{\mu}\stackrel{\neg}{\lambda}\sigma^{\sigma\rho}\lambda)
\varepsilon_{\sigma\rho\alpha\beta}+\\
&+2(\stackrel{\neg}{\lambda}\lambda)
(\stackrel{\neg}{\lambda}\sigma_{\alpha\beta}\nabla_{\mu}\lambda
-\nabla_{\mu}\stackrel{\neg}{\lambda}\sigma_{\alpha\beta}\lambda)
\end{eqnarray}
which may diverge but it may also converge to a constant value and hence the Ricci metric curvature tensor has an high-density approximation given by
\begin{eqnarray}
\nonumber
&R_{\mu\nu}=\frac{1}{2}\left(\nabla_{\mu}\stackrel{\neg}{\lambda}\nabla_{\nu}\lambda
+\nabla_{\nu}\stackrel{\neg}{\lambda}\nabla_{\mu}\lambda\right)-\\
&-\frac{1}{8}K^{\sigma\rho}_{\phantom{\sigma\rho}\mu}K_{\sigma\rho\nu}
\left(\stackrel{\neg}{\lambda}\lambda\right)
+\frac{1}{16}K_{\sigma\rho\mu}K_{\eta\zeta\nu}\varepsilon^{\sigma\rho\eta\zeta}
\left(i\stackrel{\neg}{\lambda}\gamma\lambda\right)
\end{eqnarray}
but no high- nor low-contorsional approximation may be taken any further in general; in the same limit the field equations are
\begin{eqnarray}
\nonumber
&\nabla^{2}\lambda+K^{ij\mu}\sigma_{ij}\nabla_{\mu}\lambda
+\frac{1}{2}\nabla_{\mu}K^{\alpha\beta\mu}\sigma_{\alpha\beta}\lambda-\\
&-\frac{1}{8}K^{ij\mu}K_{ij\mu}\lambda
+\frac{i}{16}K_{ij}^{\phantom{ij}\mu}K_{ab\mu}\varepsilon^{ijab}\gamma\lambda=0
\label{fieldequations}
\end{eqnarray}
that is they are the field equations we would have in the limit of masslessness.

We recall that the reason for which these fields have been defined to be neutral fields was the need of extended fields, and the fact that they lack a topological charge means that they lack a topological cause of localization. In the following we would like to empower these fields by requiring that the contorsion converges to a constant value so that the Ricci metric curvature tensor vanishes, to provide a necessary condition for the absence of the source of gravitational pull and consequently for the absence of a dynamical cause of localization: this is actually the case when the field equations are solved for fields implicitly given by the relationship
\begin{eqnarray}
\nabla_{\mu}\lambda=-iP_{\mu}\gamma\lambda
\label{fieldsolutions}
\end{eqnarray}
as plane waves; when the field is decomposed in two irreducible components if one component is given by the $\nabla_{\mu}\phi=-iP_{\mu}\phi$ plane waves then the other component is given by the $\nabla_{\mu}(\zeta\Theta\phi^{*})=\zeta\Theta(\nabla_{\mu}\phi)^{*}=\zeta\Theta(-iP_{\mu}\phi)^{*}=iP_{\mu}(\zeta\Theta\phi^{*})$ plane waves explaining why solutions that are written in terms of the plane waves have a form shown in equation (\ref{fieldsolutions}).

This means that in situations of high-density limit, we have that plane wave solutions produce the circumstance for which the energy density tends to vanish with the consequent absence of the source of gravitational attraction and thus absence of the dynamical cause of localization.

A final comment regards the fact that the solutions in plane waves of the form given by the relation $\nabla_{\mu}\lambda=-iP_{\mu}\gamma\lambda$ are such that the two irreducible components of the ELKO appear to have opposite momenta, and because we already know that these two components have different helicities then the two components have aligned spins and we conclude that repulsive effects arise between the two components of the same ELKO by the principle of exclusion.

These remarks are important for a comparison between the properties of ELKO and Dirac fields: the ELKO and Dirac fields have opposite features in the sense that as shown here the ELKO field for high-density limits has energy that tends to vanish while as shown in \cite{k} the Dirac field for the same high-density limit has energy that tend to diverge to infinity; on the other hand the ELKO and Dirac fields are also different in character as the two irreducible components of these two fields have aligned and antialigned spins respectively. 
\section*{Conclusion}
In this paper we have studied the ELKO field in the high-density limit finding plane wave solutions in such an approximation: in terms of these plane waves the contorsion tensor is constant determining the Ricci metric curvature tensor to vanish. In particular it follows from this fact that the energy density tensor is equal to zero with a consequent asymptotic freedom of gravitational sort due to the absence of the source of gravitational attraction.

This proves that ELKO fields have a dynamical behavior that determines their spreading throughout the space, and in complete contrast to the dynamical behavior of the standard Dirac fields.

\

\noindent \textbf{Acknowledgments.} I am grateful to Prof. Dharam V.~Ahluwalia for his enlightening suggestions and kind encouragement.

\


\end{document}